# Natural shaping of the cylindrically polarized beams

V. Shvedov[1], T. Fadeyeva[2], N. Shostka[2], C. Alexeyev[2] and A. Volyar[2]

[1]*Nonlinear Physics Center and Laser Physics Center, Research School of Physical Sciences and Engineering, Australian National University, Canberra ACT 0200, Australia*
[2]*Department of Physics, Taurida National University, Simferopol 95007 Crimea, Ukraine*

*Abstract . We have experimentally and theoretically shown that the circularly polarized beam bearing singly charged optical vortex propagating through a uniaxial crystal can be split after focusing into the radially and azimuthally polarized beams in vicinity of the focal area provided that the polarization handedness and the vortex topological charge have opposite signs. Quality of the polarization structure can reach unity.*

Unique properties of the cylindrically polarized beams (i.e. the radially and azimuthally polarized ones) to produce an extremely small focal spot [1] and to form the electric field with the only longitudinal component [2] under the tight focusing draw the intense attention for different applications such as the high resolution microscopy [3], particle trapping devices [4], etc. There are a lot of different ways to produce cylindrically polarized beams, e.g. intra-cavity laser devices [5], liquid crystal phase modulators [6] and others. All these devices need especially accurate alignment of special optical gadgets. The question arises: is there a simple natural way (similar to natural focusing) to produce cylindrically polarized beams without losing the beam quality? As early as in the beginning of 2000 we have observed that the polarized beams propagating along the crystal optical axis form a complex polarized structure (see, e.g., [7] and references therein), whose polarization states fill evenly all the Poincare sphere. Such a beam focused by a low aperture lens can produce two focal spots with salient polarization distributions whose structures are defined by the initial state of the beam (namely, the vortex topological charge and the spin, i.e. handedness of the polarization state), the crystal and lens parameters [8]. The question is how to transform such a complex polarization structure into the beams with the desired polarization distribution?

The aim of the paper is twofold: 1) to form experimentally and theoretically the cylindrically polarized beams via the field focusing after a uniaxial crystal and 2) to estimate quality of the polarization pattern.

1. We assume as a basis of our theoretical consideration the sketch of the experimental set-up shown in Fig.1. The circularly polarized monochromatic paraxial beam focused into a uniaxial crystal splits into two ones – the ordinary and extraordinary when propagating the crystal optical axis. The refractive indices for the beam complex amplitudes are $n_o = n_1$, $n_e = n_3^2/n_1$ where $n_1 = n_2$ and $n_3$ are the refractive indices along the major crystallographic axes [7]. The field structure of the beams after the crystal is sensitive to the signs of the spin s and the topological charge $l$ of the initial vortex beam and can be radically transformed when changing the signs of $s$ or $l$. We will denote the initial beam state as $|s\ l\rangle$. Let us consider the propagation of the paraxial vortex-beam through the optical system shown in Fig.1.

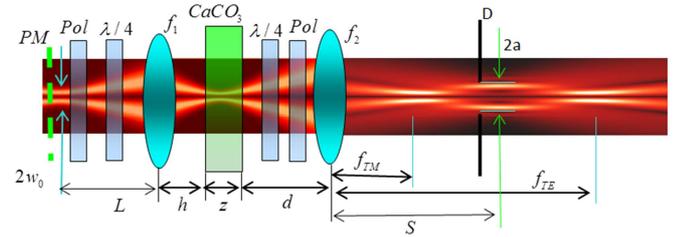

Fig.1 The sketch of the experimental set-up. The longitudinal section of the focused beam is plotted for $n_o = 1.654$, $n_3 = 1.494$ at the wavelength $\lambda = 0.634 \mu m$, L=2.5m, h=3.5cm, d=4.2cm, z=1cm, f$_1$=-5cm, f$_2$=12.5cm, $w_0 = 1mm$.

The transverse beam components can be presented in the circularly polarized basis $\{e_+, e_-\}$ for the initial beam $|1\ -1\rangle$ in the form [7]:

$$E_+ = (\Psi_o + \Psi_e)e^{-i\varphi},\ E_- = (\Psi_o - \Psi_e)e^{-i\varphi},\quad (1)$$

where $\Psi_{o,e} = (z_0 r/q_{o,e}^2)\exp\left[-ikn_1 r^2/(2q_{o,e})\right]$,
$q_{o,e} = S + f_2 q_2^{(o,e)}/(f_2 - q_2^{(o,e)})$,  $z_0 = kn_1 w_0^2/2$,
$q_1 = h + d + (L+iz_0)f_1/(f_1 + L + iz_0)$,
$q_2^{(o,e)} = q_1 + (n_1/n_{o,e})z$, $w_0$ is the radius of the initial beam at the laser's output, $k = 2\pi/\lambda$, $\lambda$ stands for the wavelength of the laser radiation, $n_1$ is the refractive index outside the crystal. In the equation (1) we made use of the ABCD rule for the centered optical system.

It is the presence of the ordinary $\Psi_o$ and extraordinary $\Psi_e$ beams with different curvature radii after the crystal that enables to form two focal spots created by the second lens. A typical longitudinal section of the beam shown in Fig.1 illustrates two clearly marked focal spots separated by the area with the dip of the beam intensity. We have specially chosen here the parameters of the optical system (e.g. the large focal length of the second lens) to show the distortion of the electric field in vicinity of the focal planes. Fig.2 demonstrates experimentally obtained intensity distributions in vicinity of the focal area. In the experiment, we used the laser diode that radiates a cylindrically symmetric beam at the wavelength $\lambda = 0.634 \mu m$. The computer-generated hologram PM produces a singly charged centered optical vortex while the polarizer *Pol* and the phase retarder $\lambda/4$ shapes the wished handedness of the circular polarization. As a uniaxial crystal we have chosen the $CaCO_3$ crystal of the length $z = 1 cm$.

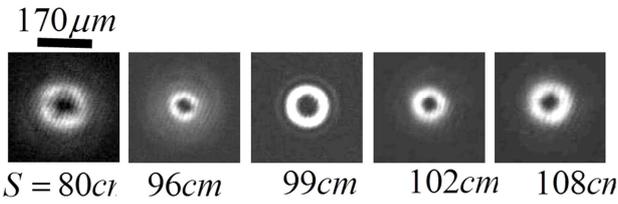

$S = 80 cm \quad 96 cm \quad 99 cm \quad 102 cm \quad 108 cm$

Fig.2 The experimentally obtained beam intensity distributions in vicinity of the focal area

All parameters of the experiment are presented in Fig.1. We can clearly observe two focuses with different polarization distributions. The computer simulation of this process is shown in Fig.3a. The field with the nearly radial polarization distribution is shaped at the distance $S = f_{TM}$. In the vicinity of the beam's axis it has the property of the transverse magnetic field (TM mode), while at the distance $S = f_{TE}$ the transverse electric field (TE mode) is shaped. The imperfections of the field distributions can be easily eliminated by changing the length of the two focal lenses. The improved maps of the field structure obtained experimentally for the optimal parameters of the system are shown in Fig.3b. For the experimental mapping of the field structure we employed the Stokes-polarimeter method described in the paper [9]

Once we change the sign of the circular polarization in the initial beam: $|1 \; -1\rangle \rightarrow |-1 \; -1\rangle$ (a simple rotation of the first polarizer axis through an angle $\pi/2$ in Fig.1) the field structure is radically transformed. The field components are described in this case as [7]:

$$E_+ = e^{-i3\varphi} \sum_{j=0}^{2} (iz_0)^{j-2} (2/j!)(r/w_0)^{2(j-2)} \left[\Psi_o / q_o^{j-2} - \Psi_e / q_e^{j-2}\right],$$
$$E_- = (\Psi_o + \Psi_e) e^{-i\varphi} \quad . (2)$$

Thus, if in the first case the $E_+$ component carries over the centered optical vortex with $l = -1$ and the $E_-$ component has the $l = 1$ vortex charge, in the second case the $E_-$ component has $l = -1$ and the triple charged vortex ($l = -3$) is embedded in the $E_+$ component. Fig.3c demonstrates the global transformation of the field structure. Typical patterns of the TE and TM mode disappear at all in this case.

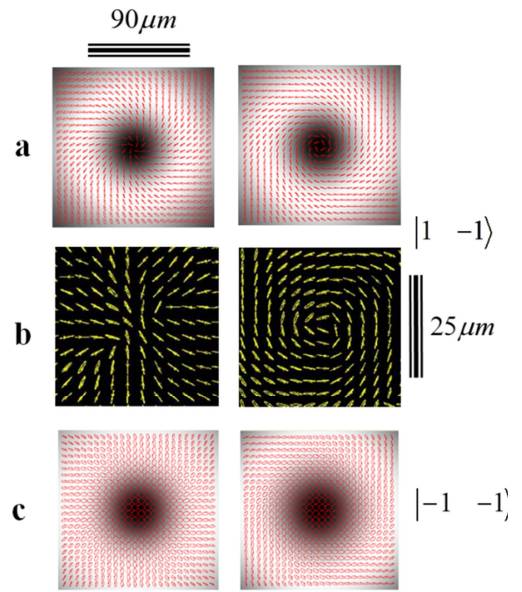

Fig.3 Polarization distributions in the focal planes $f_{TM} = 96cm$ and $f_{TE} = 102cm$ for a) $|1 \; -1\rangle$ and c) $|-1 \; -1\rangle$ states with parameters indicated in Fig.1. b) Experimental maps for the $|1 \; -1\rangle$ states with parameters of the system L=76cm, h=3cm, d=7cm, z=1cm, f$_1$=3cm, f$_2$=7cm, $w_0 = 1mm$.

It is important to notice that the two-focused state of the beam is observed only at certain parameters of the system. The criterion of the presence of two focuses is based on the following. Each partial TE and TM beams have a peak of intensity in the focal plane for the level line $r_{o,e}^2 = w_0^2 \left(1 + z^2/z_{o,e}^2\right)$ so that the peaks of the o- and e-beams are positioned at the distance $\Delta S$ from each other while the width of the level-line in the focal area is $\Delta_o \approx \Delta_e = \Delta$. Besides, each of the focal lengths must be positive. We assume that two focuses can be independently observed provided that $\Delta S > 2\Delta$:

$$\left|\text{Re}(q_2^o) - \text{Re}(q_2^e)\right| - 2\text{Im}(q_2^o) > 0,$$
$$\text{Re}(q_2^o), \text{Re}(q_2^e) > 0 \quad (3)$$

The diagram of resolubility of two focuses is shown in Fig.4. The parameters of the optical system positioned near the lane line are associated with a poor resolution of the two-focal state.

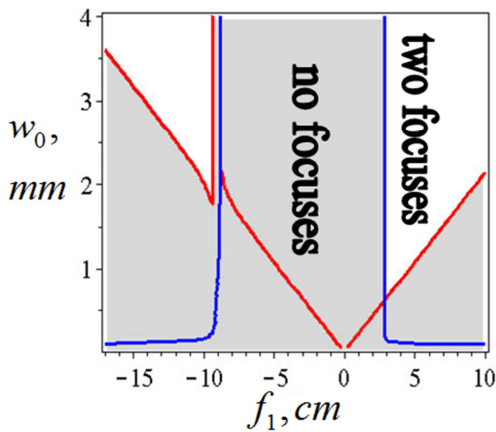

Fig.4 Diagram of the permitted and forbidden two-focus states: $L = 0, z = 1.5 cm, \quad h + d = 1 cm, \quad f_2 = 12 cm$. Color on-line

2. Distortion of the field structure in vicinity of the focal planes is caused by interference of partial TM and TE mode beams. The distortion value $\eta$ can be easily estimated as a ratio of the TE and TM mode total intensities, e.g. $\eta = \left|(I_{TM} - I_{TE})/(I_{TM} + I_{TE})\right|$. The value $\eta$ depends on the pupil radius $a$ of the diaphragm D positioned at the focal area. For example, the value $\eta \approx 0.8$ at the both focal planes for $a = 75 \mu m$ decreases up to $\eta \approx 0.6$ for $a = 120 \mu m$ if the parameters of the optical system correspond to the values in Fig.1 (the beam waist radius in the focal planes is about $w_f \approx 70 \mu m$). At the same time, the distortion quickly decreases along with the decrease of the focal length $f_2$ of the second lens so that $\eta \approx 0.99$ provided that $f_2 = 6 cm$ and $a = 40 \mu m$ (equal to the beam waist $w_f$). However, the value $\eta$ cannot be experimentally measured at least with the help of ordinary optical instruments. On the other hand, the correlation coefficient of the TE and TM modes is of the experimentally measurable value. Indeed, the key process responsible for shaping the field distribution in the crystal is the spin-orbit coupling. Any change in the spin angular momentum $S_z$ inevitably entails the change of the orbital angular momentum $L_z$ so that $S_z + L_z = const$ [10]. After the crystal, the values $S_z$ and $L_z$ are constants provided that the beam is not truncated by the diaphragm. We have recently shown [11] that the spin angular momentum is proportional to the correlation of the o- and e-beams. Thus, the value $S_z(a)$ as a function of the diaphragm radius $a$ in the truncated beam can serve as an experimentally measurable coefficient of the field structure quality:

$$S_z(a) = [I_+(a) - I_-(a)]/J(a) = 4\pi \int_0^a r \text{Re}(\Psi_o \Psi_e^*) dr / J(a), \quad (4)$$

where $I_\pm(a)$ stand for intensities of the right- and left-hand polarized components, $J(a)$ is intensity of the truncated beam. The theoretical and experimental dependencies $S_z(a)$ are shown in Fig.5. When the diaphragm radius increases the curve $S_z(S,a)$ gets flattened and rises above the line $S_z = 0$. The field structure of the beam is distorted. The fact is that the spin angular momentum of the TE and TM modes is zero. Any deviation from that value leads to the polarization structure distortion.

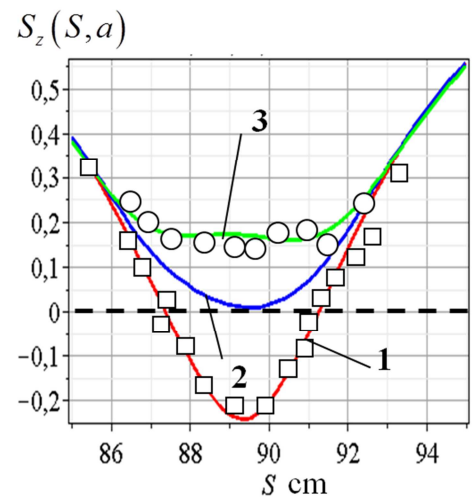

Fig.5 The beam quality $S_z(S,a)$: 1) $a = 75 \mu m$, 2) $a = 90 \mu m$, 3) $a = 120 \mu m$.
Solid lines – theory, ○, □ - experiment.

The value $\xi(S=f_{TE}, S=f_{TM})=1-|S_z|$ describes the field structure quality in terms of correlation of the TE and TM modes. The curve forms in Fig.5 for $a<75\,\mu m$ experience very weak changes so that the diaphragm radius $a=75\,\mu m$ is optimal. Notably that the field structure quality $\xi$ for the experimental patterns shown in Fig.3b is about $\xi\approx0.96$ for the diaphragm radius $a\approx30\,\mu m$ while the beam waist is about $w_f\approx25\,\mu m$.

Thus, the focusing of the crystal-propagating beam excited by the initial vortex-beam with opposite signs of the topological charge and spin can shape the radially and azimuthally polarized beams with high field structure quality in the vicinity of the two-focal region.

This work has been partially supported by the Australian Research Council